\documentclass[journal, draftcls, one column, 12pt]{IEEEtran}

\usepackage{amsmath}
\usepackage{graphicx}
\usepackage{amssymb}
\usepackage{cases}
\usepackage{color}
\usepackage{cite}
\usepackage[ruled,vlined,lined,ruled,linesnumbered]{algorithm2e}
\usepackage{relsize}
\usepackage[nodisplayskipstretch]{setspace}
\usepackage[top=1in, bottom=1in, left=0.7in, right=0.7in]{geometry}

\begin{document}
\title{Wireless Powered Communication Networks: An Overview}
\author{Suzhi~Bi, Yong~Zeng, and Rui~Zhang
\thanks{S.~Bi is with the College of Information Engineering, Shenzhen University, Shenzhen, China.}
\thanks{Y.~Zeng and R.~Zhang are with the  Department of Electrical and Computer Engineering, National University of Singapore, Singapore.}
\thanks{This work is supported in part by the National Natural Science Foundation of China (Project number 61501303).}
}

\maketitle

\vspace{-1.8cm}

\section*{Abstract}
Wireless powered communication network (WPCN) is a new networking paradigm where the battery of wireless communication devices can be remotely replenished by means of microwave wireless power transfer (WPT) technology. WPCN eliminates the need of frequent manual battery replacement/recharging, and thus significantly improves the performance over conventional battery-powered communication networks in many aspects, such as higher throughput, longer device lifetime, and lower network operating cost. However, the design and future application of WPCN is essentially challenged by the low WPT efficiency over long distance and the complex nature of joint wireless information and power transfer within the same network. In this article, we provide an overview of the key networking structures and performance enhancing techniques to build an efficient WPCN. Besides, we point out new and challenging future research directions for WPCN.

\section{Introduction}
The recent advance of microwave wireless power transfer (WPT) technology enables to build wireless powered communication networks (WPCNs), where wireless devices (WDs) are powered over the air by dedicated wireless power transmitters for communications \cite{2015:Bi,2015:Lu,2014:Krikidis}.\footnote{Another appealing line of related research is simultaneous wireless information and power transfer (SWIPT), which explores a dual use of microwave signals to transfer information jointly with energy using the same waveform (see \cite{2015:Bi,2015:Lu,2014:Krikidis} and the references therein).} Compared to conventional battery-powered networks, WPCN eliminates the need of manual battery replacement/recharging, which can effectively reduce the operational cost and enhance communication performance. Besides, WPCN has full control over its power transfer, where the transmit power, waveforms, and occupied time/frequency dimensions, etc., are all tunable for providing stable energy supply under different physical conditions and service requirements. This is in vivid contrast to \emph{energy harvesting} (EH) based approaches, where WDs opportunistically harness renewable energy in environment not dedicated to power the WDs, e.g., solar power and ambient radio frequency (RF) transmission. Because the availability and strength of renewable energy sources are mostly random and time varying, stable and on-demand energy supply to WDs is often not achievable with EH-based methods. These evident advantages of WPT over conventional energy supply methods make WPCN a promising new paradigm to the design and implementation of future wireless communication systems with stable and self-sustainable power supplies.

\begin{figure}
\centering
  \begin{center}
    \includegraphics[width=0.7\textwidth]{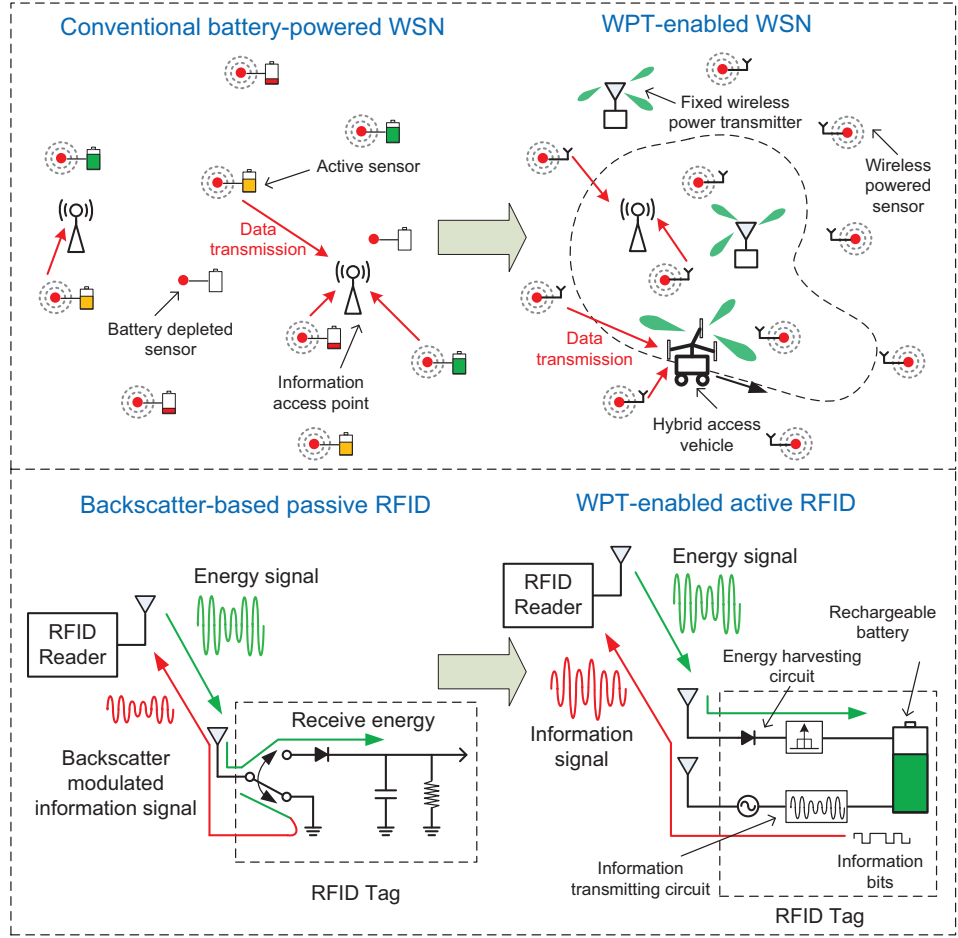}
  \end{center}
  \caption{Example applications of WPCN to conventional WSN and RFID systems. The red lines denote information flows, and the green ones denote energy flows.}
  \label{71}
\end{figure}

Current WPT technology can effectively transfer tens of microwatts RF power to WDs from a distance of more than $10$ meters,\footnote{Please refer to the website of Powercast Corp. (http://www.powercastco.com) for detailed product specifications.} while there is still significant room for improving the magnitude and range with future advancement in WPT. This makes WPCN potentially suitable for a variety of low-power applications with device operating power up to several milliwatts, such as wireless sensor networks (WSNs) and radio frequency identification (RFID) networks. Currently, commercial WPT-enabled sensors and RFID tags are already in the market. In the future, the extensive applications of WPT-enabled devices may fundamentally reshape the landscape of related industries, such as Internet-of-things (IoT) and machine-to-machine (M2M) communications. As illustrated in Fig.~\ref{71}, without the need to replace energy-depleted sensors in conventional WSN, a WPT-enabled WSN can achieve uninterrupted operation with massive number of sensors powered by fixed energy transmitters and/or a vehicle moving in a planned route used for both wireless charging and data collection. Besides, thanks to the more ample power supply from WPT, RFID devices can now expect much longer operating lifetime and afford to transmit actively at a much larger data rate and from a longer distance than conventional backscatter-based RFID communications.

Despite the potential performance improvement brought by WPCN, building efficient WPCNs is a challenging problem in practice. On one hand, the received energy level can be very low at WDs located far away from energy transmitters due to significant attenuation of microwave power over distance. This energy near-far effect can cause severe performance unfairness among WDs in different locations \cite{2014:Ju}. On the other hand, joint design of wireless energy and information transmissions is required in WPCN. First, wireless energy and information transmissions are often related, e.g., a WD needs to harvest enough energy by means of WPT before transmitting data. Second, energy transfer may share the common spectrum with communication channel, which can cause co-channel interference to concurrent information transmission. Due to the above reasons, novel physical-layer transmission techniques as well as networking protocols need to be devised to optimize the performance of WPCNs.

To tackle the above technical challenges, we provide an overview in this article on state-of-the-art techniques to build an efficient WPCN. Specifically, we first introduce the basic components and network models of WPCN. Then, we present the key performance enhancing techniques for WPCN based on the introduced system models. At last, we discuss the extensions and future research directions for WPCN and conclude the article.

\begin{figure}
\centering
  \begin{center}
    \includegraphics[width=0.8\textwidth]{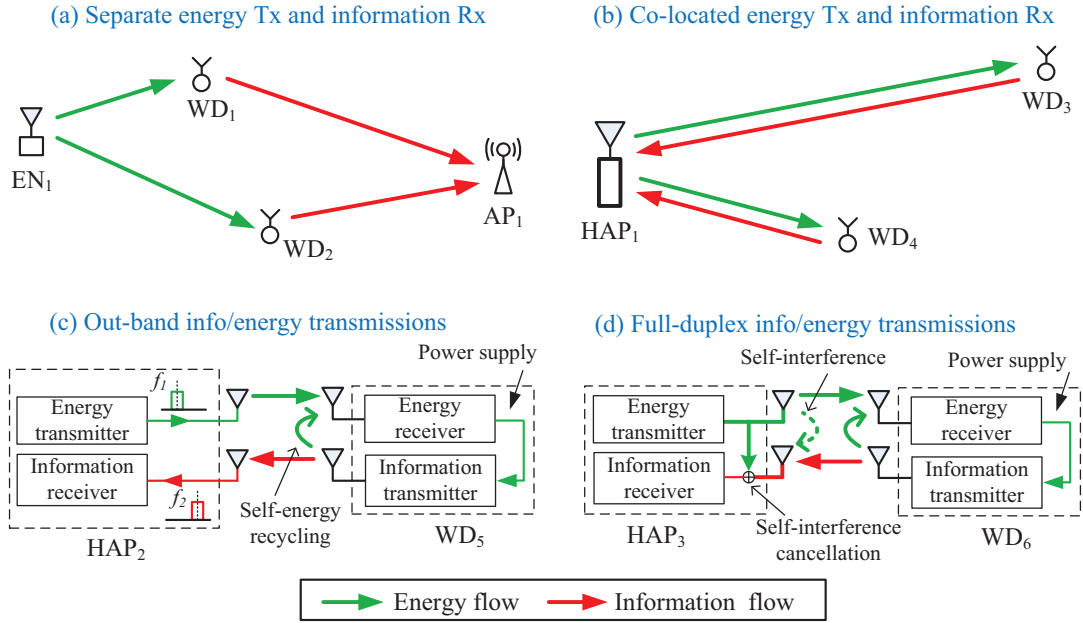}
  \end{center}
  \caption{Schematics of some basic transmitter (Tx) and receiver (Rx) models of WPCN.}
  \label{72}
\end{figure}

\section{Basic Models of WPCN}
We present in Fig.~\ref{72} some basic building blocks of WPCN. In a WPCN, energy nodes (ENs) transmit wireless energy to WDs in the downlink, and the WDs use the harvested energy to transmit their own data to information access points (APs) in the uplink. As shown in Fig.~\ref{72}(a), the ENs and APs are in general \emph{separately} located \cite{2014:Huang}, but can also be grouped into pairs and each pair of EN and AP are \emph{co-located} and integrated as a hybrid access point (HAP) as in Fig.~\ref{72}(b) \cite{2014:Ju}. The integrated HAP makes the coordination of information and energy transmissions in the network easier as compared to separated EN and AP, and also helps save the production and operation cost by sharing their communication and signal processing modules. However, it also induces a practical design challenge named ``\emph{doubly-near-far}" problem \cite{2014:Ju}, where user that is far away from its associated HAP (e.g., WD$_3$ in Fig.~\ref{72}(b)) harvests lower wireless energy in the downlink but consumes more to transmit data in the uplink than that of a user nearer to the HAP (WD$_4$). As a result, unfair user performance may occur since a far user's throughput can be much smaller than a nearby user. This user unfairness problem can be alleviated in a WPCN with separated ENs and APs. As shown in Fig.~\ref{72}(a), WD$_2$ harvests less energy than WD$_1$ because of its larger distance to EN$_1$, but also consumes less on data transmission due to its smaller distance to AP$_1$.

Furthermore, the circuit structures for energy and information transmissions are rather different. For instance, a typical information receiver can operate with a sensitivity of $-60$ dBm receive signal power, while an energy receiver needs up to $-10$ dBm signal power \cite{2015:Bi}. To maximize their respective operating efficiency, energy and information transceivers normally require different antenna and RF systems. Therefore, as shown in Fig.~\ref{72}(c) and (d), a practical WPT-enabled WD has two antenna systems, one for harvesting energy and the other for transmitting information. Similarly, an HAP with co-located energy transmitter and information receiver also needs two sets of antenna systems.

Energy and information transmissions can be performed either in an \emph{out-band} or \emph{in-band} manner. As shown in Fig.~\ref{72}(c), the out-band approach transfers information and energy on different frequency bands to avoid interference. However, energy transmission in general needs to use pseudo-random energy signal, which occupies non-negligible bandwidth, to satisfy the equivalent isotropically radiated power (EIRP) requirement on its operating frequency band imposed by radio spectrum regulators such as FCC (Federal Communications Commission). To enhance the spectrum efficiency, in-band approach allows the information and energy to be transmitted over the same band. In this case, however, energy transmitters may cause co-channel interference at information receivers, especially when an energy transmitter and an information receiver co-locate at an HAP that may receive strong self-interference. A practical  solution is to separate energy and information transmissions in different time slots \cite{2014:Ju}, which, however, reduces the time for information transmission and thus the system throughput. A point to notice is that a WD can in fact operate in an information/energy full-duplex manner, which is able to transmit information and harvest energy to/from the AP/EN (HAP) in the same band. For instance, when EN$_1$ and AP$_1$ in Fig.~\ref{72}(a) are well separated, i.e., energy transmission does not cause strong interference to information decoding, it is feasible for WD$_1$ to simultaneously receive energy from EN$_1$ and transmit information to AP$_1$. In addition, as shown in Fig. 2(c), the information/energy full-duplex operation enables an additional benefit known as \emph{self-energy recycling}, where a WD can harvest additional RF energy from its own transmitted information signal \cite{2015:Zeng1}. Evidently, a full-duplex WD can benefit from high loop-link channel gain from its information transmitting antenna to energy receiving antenna. Therefore, the receiving antenna should be placed as close to the transmitting antenna as possible, yet without disturbing its radiation pattern.

As shown in Fig.~\ref{72}(d), another promising solution for in-band approach is to use \emph{full-duplex} HAP, which is able to transmit energy and receive information to/from WDs simultaneously in the same frequency band \cite{2014:Ju1}. Notice that the full-duplex operation of HAP is different from that of WD in the sense that the energy transmission can cause severe self-interference to the information decoding. In this case, a low loop-link channel gain is practically desired to mitigate the harmful self-interference, e.g., through directional antenna design or large antenna separation. A full-duplex HAP can also perform \emph{self-interference cancelation} (SIC) to further reduce the interference power, using analog/digital SIC circuitry and hybrid signal processing approaches, etc \cite{2014:Sabharwal}. Although some recent studies suggest that perfect self-interference cancelation in wireless channel is difficult, full-duplex HAP has the potential to provide folded spectrum efficiency improvement than conventional half-duplex energy/information transmissions.

\begin{figure}
\centering
  \begin{center}
    \includegraphics[width=0.6\textwidth]{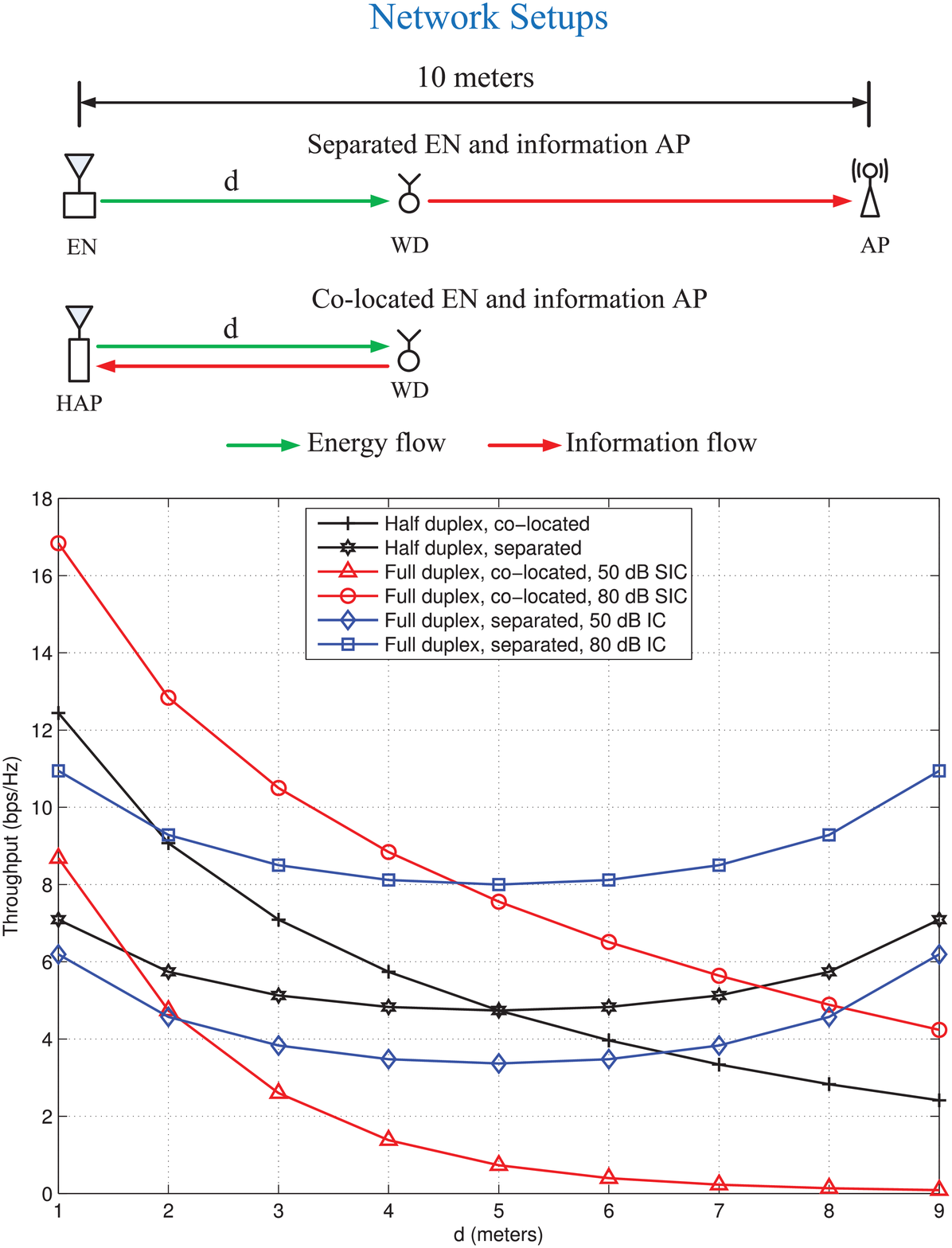}
  \end{center}
  \caption{Performance comparisons of different operating models in WPCN, where the network setups are illustrated at the top.}
  \label{73}
\end{figure}

In Fig.~\ref{73}, we present a numerical example comparing the performance of different operating models in WPCN. For the simplicity of illustration, we consider a simple WPCN consisting of only one WD, one EN and one information AP. Specifically, we consider the following six models:
\begin{itemize}
  \item half-duplex information/energy transfer using separated EN and information AP or an integrated HAP;
  \item full-duplex information/energy transfer using an HAP that can achieve either $50$ or $80$ dB SIC. That is, the interference power received by the receiving antenna is further attenuated by $50$ or $80$ dB by analog and/or digital SIC techniques before information decoding;
  \item full-duplex information/energy transfer using separated EN and information AP, where the AP can achieve either $50$ or $80$ dB interference cancellation (IC) in the received energy signal from the EN (assuming known energy signal waveform at the AP).
\end{itemize}
For half-duplex information/energy transfer, information and energy transmissions can either operate on different frequency bands (out-band) or different time slots (in-band), with the same achievable throughput performance. Here, we consider only the in-band method to avoid repetition. As shown at the top of Fig.~\ref{73}, in the case of separated EN and AP, the distance between the EN and AP is $10$ meters, and the WD locates on the line connecting them. For full-duplex operations, we assume a $-10$ dB loop-link power from the WD's information transmit antenna to its own energy harvesting antenna, such that it can achieve $10\%$ self-energy recycling. On the other hand, the loop-link power from the HAP's energy transmitting antenna to its information receiving antenna is assumed to be $-40$ dB. In addition, the transmit power of the EN is $0.1$ watt, the carrier frequency is $915$ MHz with $20$ MHz operating bandwidth, the wireless channels are assumed to follow a free space path loss model, and the receiver noise power spectrum density is $-169$ dBm/Hz.

We plot the achievable data rates of different models in bits/sec/Hz (bps/Hz) as the distance $d$ between the WD and the EN (or HAP) varies from $1$ to $9$ meters. For fair comparison between half-duplex and full-duplex operations, the time allocation ratio between energy and information transmissions is optimized for half-duplex scheme at each of the WD's locations. We can see that the data rates of using an HAP quickly degrade as the separation between the HAP and WD increases due to the doubly-near-far effect in signal attenuation. Using separated EN and AP, on the other hand, can achieve a more stable performance under distance variation because both the energy harvested and consumed decrease as the distance between the EN and WD increases. When full-duplex method is used, we can see that the data rate of full-duplex HAP with $80$ dB SIC strictly outperforms that with half-duplex HAP. However, the full-duplex HAP with $50$ dB SIC produces close-to-zero data rate even when the distance between WD and HAP is moderate, because in this case the residual self-interference overwhelms the received information signal. For full-duplex operation with separated EN and AP, the IC capability of the AP is also a critical factor that determines the communication performance. Specifically, the full-duplex operation achieves strictly higher data rate than half-duplex scheme when the AP is able to cancel $80$ dB interference, while achieving lower data rate when the IC capability of the AP is decreased to $50$ dB.

\section{Key Techniques for WPCN}
\begin{figure}
\centering
  \begin{center}
    \includegraphics[width=0.80\textwidth]{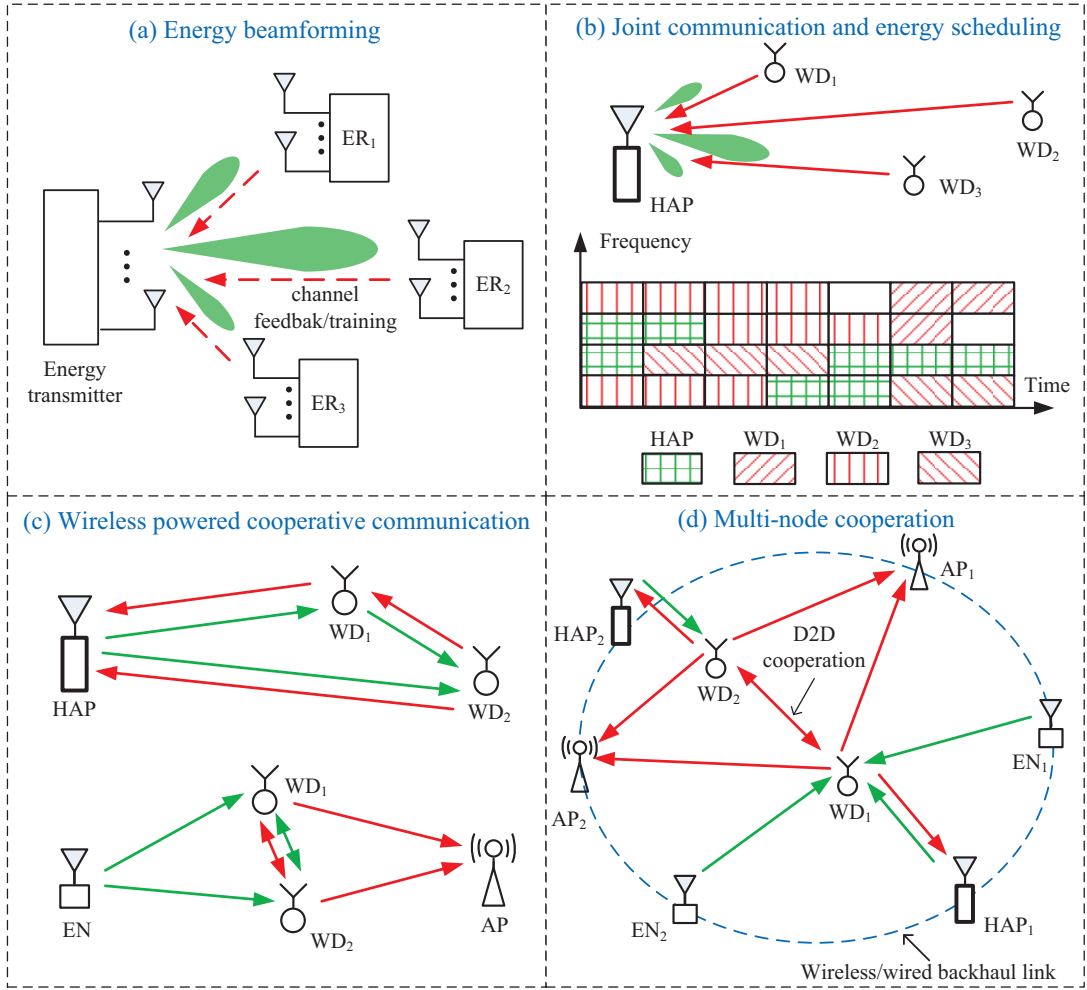}
  \end{center}
  \caption{Schematics of key performance enhancing techniques for WPCN. The green lines denote energy transfer and the red ones denote information transmission.}
  \label{74}
\end{figure}

By applying the above basic operating models, we are able to build more complex WPCNs with larger number of nodes for various different applications. In practice, the performance of WPCN is fundamentally constrained by the low efficiency and short range of WPT, and also the limited resources for both energy and information transmissions. In this section, we introduce some useful techniques to enhance the performance of WPCN. In particular, we divide our discussions into four parts: energy beamforming, joint communication and energy scheduling, wireless powered cooperative communication, and multi-node cooperation, as illustrated in Fig.~\ref{74}. The introduced methods, as well as their combined use, can effectively extend the operating range and increase the capacity of WPCN, making WPCN a viable solution to more extensive applications. For better exposition, we assume information/energy half-duplex operation in this section, while leaving the discussions on the extensions to full-duplex operation in the next section.

\subsection{Energy Beamforming}
To achieve efficient energy transfer, WPT generally requires highly directional transmission by using high-gain antennas to focus the energy in narrow energy beams towards the energy receivers (ERs). For WPT in fixed line-of-sight (LOS) links, the conventional large aperture antennas, such as dish or horn antennas, could be employed; whereas for mobile applications with dynamic channel environment, electronically steerable antenna array enabled \emph{energy beamforming} technique \cite{2015:Bi} is more suitable to flexibly and efficiently direct wireless energy to ERs by adapting to the propagation environment.

With energy beamforming, the energy signals at different antennas are carefully weighted to achieve constructive superposition at intended ERs. To maximize the received power level, the energy transmitter (ET) in general requires accurate knowledge of the channel state information (CSI), including both magnitude and phase shift from each of the transmit antennas to each receive antenna of different ERs. As shown in Fig.~\ref{74}(a), one method to obtain CSI at the ET is via forward-link (from ET to ER) training and reverse-link (from ER to ET) feedback. However, different from the conventional channel training design in wireless communication systems, where the major concern is the bandwidth/time used for transmitting training signals, the channel training design for WPT is constrained by the limited energy available at the ER to perform channel estimation and send CSI feedbacks. Intuitively, more accurate CSI knowledge can be obtained by the ET if the ER uses more energy to perform channel estimation and/or feedback. However, the energy cost to the ER may eventually offset the energy gain achieved from a more refined energy beamforming used by the ET with more accurate channel knowledge. In particular, the energy cost can be prohibitively high for ET with a large antenna array, as the channel estimation/feedback overhead increases proportionally to the number of antennas at the ET. Instead, reverse-link training method is more suitable for estimating large-array CSI. Specifically, training signals are sent in the reverse direction by the ER so that the CSI can be directly estimated at the ET without any channel estimation or feedback by the ER. In this case, the training overhead is independent of the number of antennas at the ET. However, the ER still needs to carefully design its training strategy, such as the transmit power, duration, and frequency bands, to maximize the \emph{net} harvested energy \cite{2015:Zeng}, i.e., the energy harvested at the ER less that consumed on sending training signals.

Besides the limited energy constraint at the ER, training design in WPT systems may also be constrained by the limited hardware processing capability of ER. For instance, some low-cost wireless sensors may not have adequate baseband processing units to perform conventional CSI estimation and/or feedback. To tackle this problem, new limited feedback methods need to be developed. For instance, one-bit information feedback signal can be sent from the ER to indicate the increase or decrease of the received power level during each training interval as compared to the previous one, based on which the ET can iteratively update its channel matrix estimation from the feedback using a cutting-plane algorithm \cite{2014:Xu1}. It is proved that this simple channel estimation method can converge to the exact channel matrix after finite number of iterations.

\subsection{Joint Communication and Energy Scheduling}
Communication and energy transfer are often related in a WPCN. On one hand, downlink energy transfer strategy is based on the energy demanded by the WDs to satisfy their uplink communication quality requirements. On the other hand, uplink information transmission is causally constrained by the amount of energy available at each WD after harvesting energy by means of WPT in the downlink. Therefore, information and energy transmissions should be jointly scheduled to avoid co-channel interference and optimize the overall system performance. As shown in Fig.~\ref{74}(b), time-frequency resource blocks in a WPCN can be allocated dynamically either to the HAP for energy transfer in the downlink or to the WDs for information transmission in the uplink, based on a joint consideration of the wireless channel conditions, battery states, communication demands and performance fairness among the WDs. For instance, to tackle the doubly-near-far problem, user fairness can be improved by allocating more resource blocks to the far user WD$_2$ and less to near user WD$_1$ in Fig. 4(b). It could also occur that no transmission is scheduled at some resource blocks because of the poor wireless channel conditions due to fading. Similar dynamic resource allocation method can also be extended to a WPCN with separated EN and AP, where the wireless channels for energy and information transmissions are independent. In practice, real-time information/energy scheduling is a challenging problem because of the time-varying wireless channels and the causal relationship between current WPT and future information transmissions.

Communication and energy scheduling can also be performed in the spatial domain when EN and AP are equipped with multiple antennas\cite{2014:Liu}. Specifically, energy beamforming technique can be used by an EN to steer stronger energy beams towards certain users to prioritize their energy demands. At the same time, \emph{spatial division multiple access} (SDMA) along with multi-user detection technique can be used by the AP to allow multiple users to transmit on the same time-frequency resource block. In this case, uplink transmit power control can be applied to balance the throughput performance among all the users. In general, SDMA is a more spectrally efficient method than time/frequency-division based multiple access methods. Besides, energy beamforming and SDMA can be combined with dynamic time-frequency resource allocation to further enhance the system performance in WPCN.

\subsection{Wireless Powered Cooperative Communication}
In addition to the above techniques, another promising approach is wireless powered cooperative communication, where users are allowed to share their resources, e.g., energy and time, and communicate to the AP collaboratively. As shown in Fig.~\ref{74}(c) with an HAP serving two users, the near user WD$_1$ with ample energy supply can use part of its energy and transmit time to help relay the data transmission of far user WD$_2$. Specifically, the relay protocol can be designed to consist of three time slots. In the first time slot, the HAP performs WPT and the users harvest energy; in the second time slot, WD$_2$ transmits its data to WD$_1$ for decoding; in the third time slot, WD$_1$ encodes WD$_2$'s message together with its own message and sends to the HAP. Evidently, WD$_2$ can benefit from the cooperation due to the shorter communication range compared to the direct communication with the HAP. Meanwhile, although WD$_1$ consumes energy and time on helping WD$_2$, its data rate loss due to cooperation can also be made up by an overall longer data transmission time. This is because the gain from user cooperation allows the HAP to allocate more time for data transmission instead of WPT \cite{2014:Ju2}. Besides communication cooperation, users can also perform peer-to-peer \emph{energy cooperation}, e.g., WD$_1$ directly transmits its excessive energy to WD$_2$. This potential win-win situation makes user cooperation an attractive and low-cost method to improve the overall efficiency of WPCN.

The application of wireless powered cooperative communication to a WPCN with separated EN and AP is also illustrated in Fig.~\ref{74}(c). Similarly, after harvesting the energy broadcasted by the EN, the users can perform communication and/or energy cooperation to improve the performance of each other. In particular, some spectrally efficient cooperative communication methods, such as distributed space-time coding, can be applied when the communication links between the two users are sufficiently reliable.

\subsection{Multi-node Cooperation}
As illustrated in Fig.~\ref{74}(d), besides the device-to-device (D2D) cooperation between WD$_1$ and WD$_2$, the multiple ENs and information APs can also cooperate for more efficient energy and information transmissions. Specifically, the ENs and information APs (including HAPs) are interconnected by wired/wireless backhaul links for exchanging user data and control signals that enable them to operate collaboratively in serving the WDs. In the downlink energy transfer, the collaborating ENs form a virtual MIMO (multiple-input multiple-output) system, which is able to perform \emph{distributed energy beamforming} to maximize the receive energy level at target WDs, e.g., EN$_1$, EN$_2$ and HAP$_1$ cooperatively transfer energy to WD$_1$. For the uplink information transmission, the collaborating APs essentially form a \emph{coordinated multi-point} (CoMP) system, which is able to jointly decode user messages from the signals received across multiple APs, e.g., AP$_1$, AP$_2$ and HAP$_1$ jointly decode the message of WD$_1$. Notice that the downlink energy transfer and uplink information transmission can be performed simultaneously on the same frequency band without causing interference, e.g., concurrent energy transfer to WD$_1$ and data transmission of WD$_2$. This is because the APs can cancel the interference from the energy transfer using the predetermined energy signals informed by the ENs. This fully centralized processing scheme can provide significant beamforming gain in the downlink energy transfer and spatial diversity/multiplexing gain in the uplink information transmission. However, its implementation can be very costly in a large-size WPCN due to the practical requirements, such as high computational complexity, large control signaling overhead, heavy backhaul traffic, and accurate multi-node synchronization. In practice, it may be preferable to use a hybrid scheme that integrates both centralized and distributed processing methods to balance between performance and implementation cost.

An important problem that directly relates to the long-term performance of a multi-node WPCN, e.g., average network throughput, is the placement of ENs and APs. When the WDs are fixed in location, e.g., a WSN with sensor (WD) locations predetermined by the sensed objects, the problem becomes determining the optimal number and locations of ENs and APs to satisfy a certain energy harvesting and communication performance requirements \cite{2015:Bi1}. The node placement problem in WPCN is different from that in conventional wireless communication networks, where only information APs are deployed. Intuitively, the high energy consumption of a WD that is far from any AP can now be replenished by means of WPT via deploying an EN close to the WD. In general, the placements of ENs and APs should be jointly optimized to enhance the performance of a WPCN, such as throughput, device operating lifetime, and deployment cost.

\section{Extensions and Future Directions}
Besides the discussions in the previous sections, WPCN also entails rich research problems of important applications yet to be studied. In this section, we highlight several interesting research topics that we deem particularly worth investigating.

\subsection{Extensions}
Energy beamforming is a key enabling technique in WPCN. As we have discussed in the previous section, efficient energy beamforming design requires accurate CSI at the energy transmitter (CSIT), which is often not available due to limited energy and/or simplified hardware at ERs. Besides the introduced reverse-link training and limited feedback methods that reduce the cost on CSIT estimation, energy beamforming design based on imperfect or statistical CSIT knowledge is also a practical yet challenging problem. The problem becomes even more challenging when we take into consideration the non-linear energy conversion efficiency of a practical energy receiver, where the conversion efficiency in general increases with the received RF signal power and degrades if the received power is above a certain threshold.

Meanwhile, the future advance in full-duplex technology is expected to provide folded performance improvement over the conventional half-duplex information/energy transfer method. For instance, a full-duplex HAP is able to transfer energy to and at the same time receive the data transmissions from the WDs on the same frequency band. As a result, the joint communication/energy scheduling design in WPCN needs to be revised, without the need to allocate orthogonal time/frequency for information and energy transmissions as in half-duplex based systems. For wireless powered cooperative communication, a full-duplex WD\footnote{Here full-duplex refers to the ability of a WD to perform self-interference cancelation, besides simultaneous energy harvesting and information transmission.} can transmit its own data to the information AP (or full-duplex HAP) while receiving concurrent energy transfer from the EN (or full-duplex HAP) and data transmission from its collaborating WD, given that the interference from both information and energy signals can be effectively canceled at the information receiver of the WD.

In addition, the two-user wireless powered cooperative communication model can be generalized to multi-user WPCN with a cluster-based structure, where a near user to the HAP can act as a relay for a cluster of users. This cluster-based structure can be very useful in a large-size WPCN with many poor direct WD-to-HAP links. In this case, the cluster-head nodes will be responsible for coordinating intra-cluster communications, relaying data traffic to/from the HAP, and exchanging the control signals with the HAP, etc. As a result, some cluster-head nodes may quickly deplete their batteries although they may actually harvest more energy than the other non-cluster-head nodes. Possible solutions include using energy beamforming to steer strong energy beams to prioritize the energy supply to cluster-head nodes, or using a hybrid structure that allows opportunistic direct WD-to-HAP communication of non-cluster-head nodes to reduce the energy consumption of the cluster-head nodes.

\begin{figure}
\centering
  \begin{center}
    \includegraphics[width=0.7\textwidth]{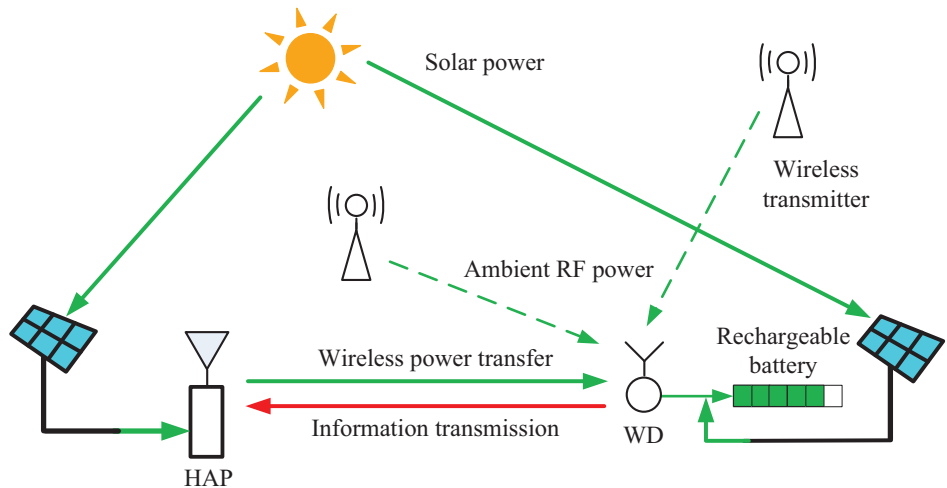}
  \end{center}
  \caption{Illustration of a WPCN with hybrid energy sources.}
  \label{75}
\end{figure}

\subsection{Green WPCN}
Energy harvesting methods can be combined with WPT to build a green and self-sustainable WPCN that requires less energy to be drawn from fixed power sources by the ENs. As illustrated in Fig.~\ref{75}, energy harvesting techniques can be applied at both the ENs and WDs. Specifically, a WD can opportunistically harvest renewable energy in the environment, such as solar power and ambient RF radiation, and store the energy in a rechargeable battery. On one hand, when the intensity of renewable energy is strong at most WDs, ENs can turn off energy transfer to avoid waste of energy due to limited battery capacity. On the other hand, conventional WPT can be used to power the WDs when effective energy harvesting is not feasible at most WDs. In between, a hybrid power method using both energy harvesting and WPT can be adopted, where ENs can perform transmit power control or use energy beamforming to concentrate transmit power to the users who harvest insufficient renewable energy. In a green WPCN with hybrid power sources, the key challenge is to achieve timely switching between different operating modes and design efficient energy transmit strategies, to minimize the energy drawn from fixed power sources while satisfying the given communication performance requirements. In general, the optimal design requires the joint consideration of a number of factors, such as the current and predicted renewable energy intensity, battery state information, and wireless channel conditions, which is still open to future investigation.

\begin{figure}
\centering
  \begin{center}
    \includegraphics[width=0.65\textwidth]{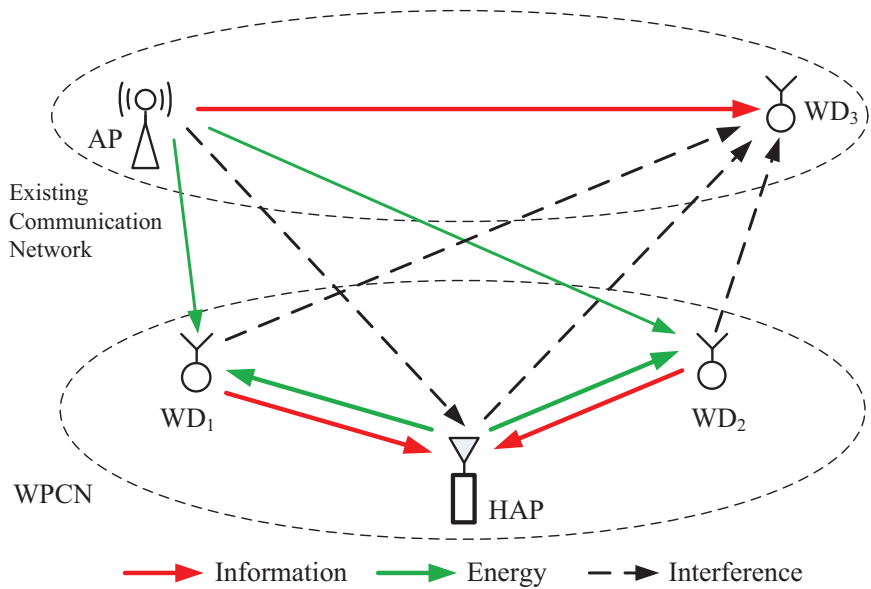}
  \end{center}
  \caption{Schematics of a WPCN under spectrum sharing with an existing communication network.}
  \label{76}
\end{figure}

\subsection{Cognitive WPCN}
In practice, a WPCN is likely to coexist with other existing communication networks, while they can cause harmful co-channel interference to each other when operating simultaneously on the same frequency band. As shown in Fig.~\ref{76}, the WPCN can cause interference to the information decoding at WD$_3$ in an existing communication network. At the same time, the transmission of the AP in the communication network can also cause interference to the information decoding at the HAP in the WPCN. Notice that, although transmission in the existing communication network produces harmful interference to the HAP's information decoding, it also provides additional energy to harvest for the users (WD$_1$ and WD$_2$) in the WPCN. Given limited operating spectrum, a WPCN should be made cognitive to efficiently share the common frequency band with existing communication networks. In particular, a cognitive WPCN can be either cooperative or noncooperative with the existing communication networks. On one hand, a cooperative WPCN protects the communication of the existing communication networks. Similar to the primary/secondary network setup in conventional cognitive radio networks, the WPCN (secondary network) designs its transmit strategy to optimize its own performance given that its transmission will not severely degrade the communication performance of the existing communication networks (primary network) \cite{2015:Lee}. On the other hand, a noncooperative WPCN designs its transmission strategy to optimize its own system performance, while only with a secondary consideration to minimize its impact to the existing communication networks \cite{2015:Xu}. In practice, some incentive schemes that can promote mutual cooperations may be a promising solution to the coexisting problem between a cognitive WPCN and conventional communication networks.

\section{Conclusions}
In this article, we have provided an overview on the basic models of WPCN and the corresponding performance-enhancing techniques to build efficient WPCNs. Compared to the battery-powered and environment energy harvesting based communications, WPCN significantly improves the throughput and reliability of the network. Although many techniques introduced for WPCN appear to be similar to those in conventional wireless communication networks, the additional dimension of energy transfer requires more sophisticated system design, and yet also brings in valuable opportunities to solve the fundamental energy-scarcity problem for wireless communications. We foresee that WPCN will be a necessary and important building block for future wireless communication systems to achieve energy self-sustainable device operations.

\end{document}